\documentclass[proceedings]{JHEP}

\title{Boundary bound states in integrable quantum field theory}
\author{E. Corrigan\thanks{E-mail: ec9@york.ac.uk}\\ Department of Mathematics, 
\\University of York, 
\\York YO10 5DD, UK\\
\ \\
{\rm and}\\
\ \\
Department of Mathematical Sciences,\\
University of Durham,\\
Durham DH1 3LE, UK\\}

\abstract{The purpose of this talk is to sketch some recent progress which has been
made in calculating non-perturbatively the reflection factors for the
sinh-Gordon model restricted to a half-line by integrable boundary conditions.
The essential idea is to calculate the energy spectrum of boundary breathers in
two independent ways; firstly by using the boundary bootstrap and secondly by
quantizing the classical solutions corresponding to boundary breathers. Comparing
these two calculations provides a way to determine the dependence of the reflection
factors on the parameters introduced into the Lagrangian by the boundary conditions.
The basic idea is illustrated using a massive free scalar field with a linear
boundary condition confining it to a half-line.}
\conference{Non-perturbative Quantum Effects 2000}
\preprint{hep-th/0010094}
\begin{document}

\section{Introduction}

Although boundary conditions are extremely important for field theory---think
of monopoles or vortices, for example, or instantons---the study of integrable
quantum field theory restricted to a half-line or  an interval by integrable
boundary conditions is not yet fully developed. In particular, many interesting
areas are yet to be explored and the foremost of these is the classification
of the reflection factors which determine the manner in which a particle rebounds
from a boundary. In 1993 Ghoshal and Zamolodchikov \cite{Ghosh94a,Ghosh94b}
ignited activity in this subject by analysing the possible behaviour of the sine-Gordon
model restricted to a half-line. They discovered a two-parameter family of boundary 
conditions which preserved integrability, in the sense of preserving after suitable
modification all the energy-like conserved quantities, while violating  those which
were momentum-like. Clearly, the same facts would be true of the sinh-Gordon model,
which is even simpler in the sense of having a particle spectrum consisting of a single
scalar particle and nothing else. The reflection factors for the states belonging to
the sine-Gordon particle spectrum (solitons, anti-solitons and breathers) were
computed although the relationship between two parameters appearing in the reflection 
factors
and the two parameters appearing in the classical Lagrangian 
specifying the boundary
conditions remained largely obscure. The purpose of developing the techniques reviewed
here was to remove this obscurity. It seems such a simple matter that it is
surprising that it has taken so many years to find a way to resolve it.

Apart from the troubles relating Lagrangian parameters to the parameters appearing
in reflection factors there are other mysteries too. For example, the sinh-Gordon model
is the simplest of the affine Toda field theories and yet it is the only one in
the $ade$ series of cases with
a set of boundary conditions with free parameters. For all the rest, the possible
integrability-preserving boundary conditions unexpectedly form a discrete set 
\cite{Cor94a, Cor95, Bow95}. It is not at all clear what the reflection 
factors are which correspond to each member of this discrete set, although a variety of
reflection factors have been found using algebraic means (see
for example  \cite{Del99a}). Our long-term goal is to find a complete classification
and a full understanding of the curious pattern of integrable boundary conditions
discovered in \cite{Bow95}.

For a review of the background to some of these questions see \cite{Cor96}.

\section{A free scalar field with a linear boundary condition}

In this section, to set the scene, consider a free massive scalar field confined
to the left half-line ($x<0$) by a linear 
boundary condition at $x=0$. Its equation of motion and boundary condition are:
\begin{equation}\label{linear}
 (\partial^2 +m^2)\phi =0, \ x<0; \ \partial_x\phi=-\lambda\phi, \
 x=0.
 \end{equation}
 Clearly, there is a set of solutions to (\ref{linear}) corresponding to a
 superposition of left and right-moving waves of the form
 \begin{equation}\label{linearsolution}
 \phi=e^{-i\omega t}\left(e^{ikx} + R_0(k) e^{-ikx}\right) + c.c.
 \end{equation}
 with $w^2=m^2+k^2$ and a `reflection factor' given by
 \begin{equation}\label{classicalR}
 R_0=\frac{ik+\lambda}{ik-\lambda}.
 \end{equation}
 If $\lambda <0$ there is also a periodic solution of the form (whose
 existence is indicated
 by a pole in $R_0$ at $k=-i\lambda$),
 \begin{equation}\label{breather}
 \phi =A\cos\omega t \, e^{-\lambda x}, \ \omega^2=m^2-\lambda^2,
 \end{equation}
 provided also $0>\lambda >-m$. Clearly, this solution is slightly delicate
 and fails to exist at all for a massless free field. It is a harmonic oscillator 
 `glued' to the boundary. Its energy is computed to be ${\cal E}$ where
 \begin{eqnarray}
 &&{\cal E}=\int_{-\infty}^0 dx \left( \dot\phi^2 + {\phi^\prime}^2 +m^2\phi^2\right)
  +
 \lambda \phi^2(0,t)\nonumber\\ &&\hspace{2cm} =-\frac{\omega^2 A^2}{2\lambda} 
 \nonumber
 \end{eqnarray}
 Quantizing this periodic system leads to a tower of states with energies 
 $(n+1/2)\omega$.
 
 On the other hand, in the context of integrable field theory, we know that the
 S-matrix for the free massive particle is unity and the reflection factor is 
 identical to the classical version (\ref{classicalR}). That is, setting $\omega
 =m\cosh\theta,\ k=m\sinh\theta$ we have the relationship between `in' and `out'
 states
 \begin{equation}
 |\theta>_{\rm out}=R_0(k)|-\theta >_{\rm in} \nonumber
 \end{equation}
 where `in' refers to a particle of rapidity $\theta$ approaching the boundary
 and `out' refers to the same particle at a later time after reflecting from the 
 boundary. The pole in $R_0$ at $k=-i\lambda$ indicates a boundary bound state at
 an imaginary rapidity ($\theta=i\psi=-i \sin^{-1}(\lambda /m)$) and the boundary 
 bootstrap
 \cite{Ghosh94a, Fring95, Cor94a}, indicates that the energy of the 
 excited
 boundary is
 given by
 \begin{equation}
 {\cal E}_1={\cal E}_0+m\cos\psi = {\cal E}_0+\omega ,
 \end{equation}
 with an associated reflection factor again equal to $R_0$. The reason for this is that
 the reflection factor $R_1$ for the particle rebounding from the excited boundary is 
 given
 by
 \begin{equation}\label{bootstrap}
 R_1(\theta)=S(\theta -i\psi)R_0(\theta)S(\theta+i\psi)
 \end{equation}
 and the S-matrix is unity. Repeating the process leads to the tower of harmonic 
 oscillator states.
 
 So, the basic idea which we wished to exploit was to follow this line of reasoning
in the sinh-Gordon model, constructing a tower of boundary states in two different
 ways: firstly, using the formal bootstrap, and secondly, using a direct quantization
 of periodic solutions corresponding to an `oscillator' attached to the boundary.
 It is expected that the same approach will be applicable to the other Toda
 theories as well.
 
 \section{The sinh-Gordon model}
 
 We begin by establishing notation and reminding ourselves of certain relevant facts 
 about the sinh-Gordon model on a half-line. Its field equation and integrable boundary
 conditions are
 \begin{eqnarray}\label{shGordon}
 &&\partial^2\phi =-\frac{\sqrt{8}m^2}{\beta^2}\sinh(\sqrt{2}\beta\phi), \ x<0
 \\
 && \partial_x\phi =\frac{\sqrt{2}m}{\beta}\left(\epsilon_0e^{-\beta\phi/\sqrt{2}}
 -\epsilon_1 e^{\beta\phi/\sqrt{2}}\right), \ x=0,\nonumber
 \end{eqnarray}
 where $m$ is a mass parameter (which we shall set to unity), $\beta$ is the bulk 
 coupling,
 and $\epsilon_0$ and $\epsilon_1$ are the two parameters introduced by the boundary 
 conditions. Notice that when $\epsilon_0\ne \epsilon_1$ the bulk symmetry
 under the transformation $\phi\rightarrow -\phi$ is lost. Although it is not 
 strictly necessary to
 assume that $|\epsilon_i|\le 1$, in what follows it will 
 often be useful to put
 \begin{equation}\label{alternativeparameters}
 \epsilon_i=\cos a_i\pi, \ i=0,1,\quad 0\le a_1\le a_0\le 1.
 \end{equation}
 The basic question is to give a complete characterization of the reflection factors
 as functions of the bulk coupling and the two boundary parameters.
 
 The sinh-Gordon model is not free but its S-matrix is well-known and may be written 
 compactly as follows \cite{Fadd78}
 \begin{equation}\label{Smatrix}
 S(\theta_{12})= -\frac{1}{(B)(2-B)},\ B(\beta)=\frac{\beta^2/2\pi}{1+\beta^2/4\pi},
 \end{equation}
 using the block notation,   
 \begin{equation}
 (x)=\frac{\sinh\left(\theta_{12}/2 +i\pi x/4\right)}{\sinh\left(\theta_{12}/2 -
 i\pi x/4\right)}, \nonumber
 \end{equation}
 introduced in \cite{Bra90}.
 The S-matrix is invariant under the transformation $\beta\rightarrow 4\pi/\beta$.
 
 The reflection factors may be deduced from those computed by Ghoshal \cite{Ghosh94b}
 for the 
 sine-Gordon model, and have the form
 \begin{equation}\label{shGordonR}
 R(\theta)=\frac{(1)(1+B/2)(2-B/2)}{(1-E)(1+E)(1-F)(1+F)},
 \end{equation}
 where $E,F$ are two functions  which are independent of the rapidity but depend 
 upon the  bulk coupling and the boundary coupling  parameters in a manner which needs 
 to be determined. The classical limit 
  of (\ref{shGordonR}) as $\beta\rightarrow 0$ has been calculated 
 independently \cite{Cor95} and implies $E\rightarrow a_0+a_1,\ F\rightarrow a_0-a_1$,
 using the notation (\ref{alternativeparameters}). Indeed, (\ref{shGordonR})
 is the simplest guess having the appropriate classical limit and satisfying the
 `crossing unitarity relation' \cite{Ghosh94a, Fring94}
 \begin{equation}
 R(\theta+i\pi/2)R(\theta-i\pi/2) S(2\theta)=1.
 \end{equation}
 We are also intrigued to know what happens to the weak-strong coupling invariance 
 enjoyed by the S-matrix factor. 
 
 The expression (\ref{shGordonR}) has also been
 checked perturbatively in various ways \cite{Kim95, Cor98, Top97,Chen00} 
 but the details will not be described here.
 
 \section{Using the boundary bootstrap}
 
 Examining the expression (\ref{shGordonR}) reveals that it may have poles at 
 $\theta=i\psi, \ 0<\psi <\pi/2$ and that these poles cannot come from the numerator 
 factors, only from zeroes of the denominator. Moreover, the existence of suitable
 poles depends on the value of $E$ or $F$. For example, with $1<E<2$ there are poles
 at $\theta=i(E-1)\pi/2$. What is more, knowing the classical limit of $E$ and $F$
 is useful because it informs us that it is plausible, given the ranges of $a_0$ and 
 $a_1$, that $E$ and $F$ do not lie simultaneously in the range $(1,2)$. This fact
 is true for the classical limit, and the order $\beta^2$ perturbative calculations
 do not upset it. So, we shall proceed as though it were true and see that
 the resulting expressions for $E$ and $F$ are consistent with it. For the symmetric
 case, $(a_0=a_1$), $F=0$, and the question does not even arise. More details of these
 calculations may be found in \cite{Cor99} and \cite{Cor00}.
 
 Then, provided $1<E<2$ we may compute a tower of states using the  boundary
 bootstrap (\ref{bootstrap}). The result is a sequence of states whose energies are 
 given by
 \begin{equation}\label{energyspectrum}
 {\cal E}_{n+1}={\cal E}_n + m\cos\frac{\pi}{2}(nB-E+1),
 \end{equation}
 and for which the associated reflection factors have the form
 \begin{eqnarray}\label{otherR}
&& R_n(\theta)=R_D \frac{(1+E+B)(1-E-B)}{(1-E+nB)(1+E-nB)}\nonumber\\
&&  \times \frac{1}{(1-E+(n-1)B)(1+E-(n-1)B)},\nonumber\\
&&
 \end{eqnarray}
 where
 \begin{equation}
 R_D=\frac{(1)(1+B/2)(2-B/2)}{(1-F)(1+F)}\nonumber
 \end{equation}
 is $n$-independent and never contains contributing poles.
 
 Notice that the number of new types of pole is limited because, even if $E$ is
 initially within the appropriate range, eventually $E-1-nB$ will move out of 
 range. It
 seems natural to interpret the pole at $i(E-1 -(n-1)B)\pi/2$ as a `crossed' process
 in which the state with energy ${\cal E}_n$ drops down to the state at energy 
 ${\cal E}_{n-1}$.\footnote{
 However, there is a bit of a mystery, or at least an inconsistency of interpretation
 between this case and the free field case. Here, we have in mind that there ought
 to be a finite number of boundary states just as there are a finite number of 
 sine-Gordon  breathers at most,  at a given bulk 
 coupling (see \cite{Raj82} for a review); only in the classical
 limit do we expect to have infinitely many bound states. On the other hand, 
 it seems natural
 to have the infinite tower when we are discussing the free field boundary states. 
 Thanks to Patrick Dorey for a discussion on this issue.} If we interpreted this pole as a new state
 then the reflection factor corresponding to it would contain higher order poles
 and these in turn would need to be explained.
 
 The next part of the argument requires a set of classical solutions to the
 system of equations (\ref{shGordon}).
 
 \section{Classical boundary breathers}
 
 The classical periodic solutions which we shall use may all be described within
 Hirota's ansatz. That is, for our purposes it is convenient to
 set
 \begin{equation}\label{Hirota}
 \phi = - \frac{\sqrt{2}}{\beta}\ln \frac{\tau_0}{\tau_1},
 \end{equation}
 where
 \begin{eqnarray}\label{exponentials}
 && \tau_a = 1 +(-)^a(E_1+E_2+E_3) +A_{12}E_1E_2\nonumber \\
 &&\ \ + A_{13}E_1E_3 + A_{23}E_2E_3 +A_{12}A_{13}A_{23} E_1E_2E_3,\nonumber\\
 \end{eqnarray}
 and
 \begin{eqnarray}
 &&\hspace{.8cm} E_p=e^{a_p x +b_p t +c_p}\nonumber\\
 && \ a_p =2\cosh\rho_p, \ b_p=2\sinh\rho_p\nonumber\\
&& \hspace{.5cm} A_{pq}=\tanh^2\left(\frac{\rho_p-\rho_q}{2}\right).
 \end{eqnarray}
At this stage, the constants $c_\rho$ are not determined. 

On the
whole line, these solutions are not particularly interesting because
for each $t$ they develop a singularity at some $x$. Moreover, owing
to the singularities their total  energy will be infinite. However,
we are interested in periodic solutions defined only on the region $x<0$
and therefore it is possible that  solutions satisfying the boundary
conditions can be non-singular except at points beyond the boundary; in
effect, the singularities are hidden away in the region $x>0$.
This hope turns out to be realised rather neatly and the details are given
more fully in \cite{Cor99}. The trick is to select $E_1$ and $E_2$ to be
time dependent and periodic (ie, inevitably complex), and to satisfy $E_1^*=E_2$,
while taking $E_3$ to be real. Then, the solution given via (\ref{Hirota})
is real and  the boundary conditions can be satisfied. 

For the particular case 
where $\epsilon_0=\epsilon_1\equiv\epsilon$, it is enough to set $E_3=0$ 
and we have,
\begin{eqnarray}\label{easybreather}
&&\tau_a=1+\frac{(-)^a}{\tan\rho}\, e^{2x\cos\rho}\cos(2t\sin\rho)
\sqrt{\frac{\epsilon+\cos\rho}{\epsilon-\cos\rho}}\nonumber\\
&&\hspace{1cm} -\, e^{4x\cos\rho}\, \frac{\epsilon+\cos\rho}{\epsilon-\cos\rho}.
\end{eqnarray}
The expression (\ref{easybreather}) is clearly periodic, with period $\pi/\sin\rho$.
However, for the corresponding solution constructed via (\ref{Hirota}) to be 
non-singular, $\epsilon$ and $\rho$ must be constrained as
follows:
\begin{equation}\label{conditions}
-1 <\epsilon <0, \ \ \cos\rho \le -\epsilon .
\end{equation}
As expected, $\epsilon$ should be negative (meaning that there is a 
competition between
the different signs of the boundary and bulk contributions to the 
energy), but not too negative 
(cf the free field case). More of a surprise is the constraint on 
$\rho$ since it
means that there is a minimum frequency for the breather but, at that minimum frequency
the amplitude of the breather has shrunk to zero. This behaviour is different to the
behaviour of the sine-Gordon breathers since for those the frequency drops to zero as
the amplitude collapses to zero. It seems the boundary breathers are still breathing
faintly despite their collapse when $\cos\rho=-\epsilon$. This behaviour is
reminiscent of a standard oscillator whose amplitude may be tuned to zero independently
of its frequency. In a nonlinear system for a given boundary condition the frequency 
governs the amplitude of the periodic solution.

In the general case, the solutions are substantially more complicated. The
boundary conditions
require
\begin{eqnarray}\label{generalbreather}
&&e^{c_3}=\frac{r}{q^2},\ 
e^{c_1}=e^{c_2}=\frac{s}{\tan\rho},\  r=\frac{q_-}{q_+}\nonumber\\
&&\hspace{.2cm} s^2=\frac{1}{q^2}\, \frac{1-q^2q_+^2}{1-q^{-2}q_+^2}\, 
\frac{1-q^{-2}q_-^2}{1-q^2q_-^2}
\end{eqnarray} 
where
\begin{equation}
q=\tan\frac{\rho}{2}, \ q_\pm = \tan\frac{\pi}{4}(a_0\pm a_1).
\end{equation}
The general boundary breathers have no singularities in the region $x<0$
provided
\begin{eqnarray}\label{restrictions}
&&\cos\frac{\pi}{2}(a_0+a_1)<0, \ \cos\frac{\pi}{2}(a_0-a_1) >0\nonumber\\
&&\hspace{.5cm} 0<\cos\rho<-\cos\frac{\pi}{2}(a_0+a_1). 
\end{eqnarray}
The restrictions (\ref{restrictions}) were found numerically using Maple
assuming that the parameters $a_0$ and $a_1$ lie in the range indicated in
(\ref{alternativeparameters}). The energy of the general periodic solution
given by the data (\ref{generalbreather}) is given by \cite{Cor99}
\begin{equation}
{\cal E}_b=-\frac{8}{\beta^2}\left( 1+\cos\rho -\frac{1}{2}(\sin\frac{\pi a_0}{2}+
\sin\frac{\pi a_1}{2})^2\right).
\end{equation}

\section{Semi-classical quantization}

Now that we have a collection of periodic solutions the next step is to
quantize them to determine their energy spectrum. We shall do this by
adapting the WKB methods proposed long ago by Dashen, Hasslacher and Neveu
\cite{DHN75}. Here it will be enough to outline the prescription and more
of the details are to be found in \cite{Cor99,Cor00}. The procedure
is certainly non-perturbative but we have no way of telling at present if it 
is also exact. In the case of the sine-Gordon model in the bulk, the WKB
methods do give exact results in agreement with calculations made within the
eight vertex model (for example, see \cite{Lut80}). It remains to be seen
whether this feature will persist in the presence of boundaries. As we shall
see, there are several surprises which render the analysis for the general
periodic solution manageable.

The first ingredient we need is the classical action integrate over a single 
period of the boundary breathers. Since we know the energy already, it will be 
necessary to calculate the appropriate integral of $\dot\phi^2$. Thus,
\begin{eqnarray}\label{action}
S_{\rm class}&&\! =\int_0^T dt \int _{-\infty}^0 dx\,  \dot\phi^2 \, - T {\cal E}_b
\nonumber\\
&&\! =\frac{8\pi}{\beta^2}\left(\rho -\pi(1-\frac{a_0+a_1}{2})\right)
-T{\cal E}_b.\nonumber\\
&&
\end{eqnarray}
This is a remarkable result because the first part of the expression depends
only on the combination $a_0+a_1$. For the symmetric case $a_0=a_1$, the
expression for the action is relatively straightforward to compute analytically
by a suitable sequence of changes of integration variables. However, in the
general case the result was deduced numerically and the rather convincing numerical
evidence we have available is summarised in the appendix to \cite{Cor00}. It
would be nice to have an analytic derivation of (\ref{action}), however it
has eluded us so far.

Besides the classical action we shall also need the linear perturbations around
the  breathers regarded as a background. In other words put $\phi=\phi_0
+\eta$, where $\phi_0$ is a breather and solve the following linear
equations for $\eta$:
\begin{eqnarray}
&&\partial_t^2\eta -\partial_x^2\eta +4\eta \cosh( \sqrt{2}\beta \phi_0) =0,\ x<0
\nonumber\\
&&\partial_x\eta + \left(\epsilon_0 e^{-\beta\phi_0/\sqrt{2}} +
\epsilon_1 e^{\beta\phi_0/\sqrt{2}}\right)=0,\ x=0.\nonumber\\
&&
\end{eqnarray}
The solutions in which we are interested are asymptotically plane-wave solutions, 
which means that as $x\rightarrow -\infty$ they have the form,
\begin{equation}\label{asymptoticeta}
\eta \sim e^{-i\omega t}\left(e^{ikx}+R(k)e^{-ikx}\right),\ \omega^2=k^2+4,
\end{equation}
where $R$ is the classical reflection factor in the breather background.
Actually, we shall need the classical reflection factor in the classical
ground state background as well (this particular reflection factor, incidentally, 
is the classical limit
of Ghoshal's formula (\ref{shGordonR}). Fortunately, a further use of Hirota's
method, adding two infinitesimal exponentials to (\ref{exponentials}) in the
standard manner, allows a straightforward derivation of the reflection factors we
need. They are,
\begin{eqnarray} \label{classicalfactors}
&&R_{\rm breather}=\left(\frac{ik-2\cos\rho}{ik+2\cos\rho}\right)^2\, 
\frac{ik-2}{ik+2}\,\nonumber\\
&&\hspace{2cm}\times \frac{ik-2\cos a_+}{ik+2\cos a_+}\, 
\frac{ik+2\cos a_-}{ik-2\cos a_-}\nonumber\\
&& R_{\rm ground}=\frac{ik-2}{ik+2}\,
 \frac{ik+2\cos a_+}{ik-2\cos a_+}\, 
\frac{ik+2\cos a_-}{ik-2\cos a_-}\nonumber\\
\end{eqnarray}
and, for notational convenience it is useful to define $a_\pm =\pi (a_0\pm a_1)/2$. 
The second of the reflection
factors given in (\ref{classicalfactors}) allows us to deduce the classical limit
for $E$ or $F$ quoted earlier. Another surprising feature which we did not expect
is that the ratio of the two classical reflection factors does not depend on
the combination $a_0-a_1$. 

The linear perturbations of the breathers do not have the same periodicity as
the breathers themselves. Rather, their frequencies provide the modification to 
the action which is to be used in the WKB aproximation. Thus, we need to calculate
a quantity $\Delta$ defined by
\begin{equation}
\Delta= \frac{T}{2}\sum \left(\omega_{\rm \, breather} -\omega_{\rm ground}\right),
\end{equation}
and, in terms of this the quantum action is defined to be 
\begin{equation}
S_{\rm qu}=S_{\rm class}-\Delta .
\end{equation}
The evaluation of $\Delta$ is not quite straightforward although we are 
fortunate to have
the old work of Dashen, Hasslacher and Neveu as a guide. The method we adopted runs 
as follows. First discretize the sum over the frequencies and then regulate it by
removing an infinite piece. It is convenient to use a Dirichlet boundary condition
at $x=-L$, so that $\eta (-L,t)=0$, effectively placing the system in a box
without changing the boundary condition at $x=0$. Subsequently, the limit
$L\rightarrow\infty$ will be taken at the end of the calculation. 
Removing the infinite part
of the discrete sum corresponds to making a choice of normal-ordering in
a perturbative approach to the field theory. 

Once the
Dirichlet condition is imposed the possible discrete values of the
parameters $k$ appearing in (\ref{asymptoticeta}) are given by
\begin{equation}\label{kconditions}
e^{-2iLk_B}=R_B(k_B), \ e^{-2iLk_0}=R_0(k_0),
\end{equation}
where the subscripts $B$ and $0$ refer to the breather and ground state, 
respectively. Typically, (\ref{kconditions}) have infinitely many
real solutions together with a small number of purely imaginary solutions. The
latter do play a r\^ole which will not be discussed here. The details may be found in
\cite{Cor00}. For large $L$, it is useful to note that the difference
between $k_B$ and $k_0$ is small, prompting us to write for the solutions to
(\ref{kconditions})
\begin{equation}
(k_B)_n=(k_0)_n +\frac{1}{L}\kappa((k_0)_n),
\end{equation}
where the function $\kappa$ satisfies
\begin{equation}
e^{2i\kappa(k)}=\left( \frac{ik+2\cos\rho}{ik-2\cos\rho}\, 
\frac{ik+2\cos a_+}{ik-2\cos a_+}\right)^2.
\end{equation}
Again, we notice the curious and surprising fact that only the combination 
$a_0+a_1$ appears.

The detailed calculation of $\Delta$ may be found in \cite{Cor99, Cor00}; here
it is enough to quote the result which is pleasingly simple:
\begin{eqnarray}
&&\Delta = \pi - \frac{2}{\sin\rho}\left(\cos\rho +\cos a_+ +\rho\sin\rho
\right.\nonumber\\
&&\hspace{2.2cm}\left. +(a_+-\pi/2) \sin a_+ \right).
\end{eqnarray}

The WKB prescription instructs us to define ${\cal E}_{\rm qu}=-\partial S_{\rm qu}/
\partial T$, and then set 
\begin{equation}
S_{\rm qu} + T {\cal E}_{\rm qu}=2n\pi
\end{equation}
where $n$ is a (positive) integer, or zero. In other words, assembling all the
ingredients, we have the quantization rule
\begin{equation}
2n\pi = \frac{4}{B}(\rho -\pi /2) +\frac{8\pi}{\beta^2}(a_+-\pi /2),
\end{equation}
from which we deduce a set of angles $\rho_n$, and thence a tower of energies 
${\cal E}_n$ satisfying
\begin{eqnarray}\label{WKBenergyspectrum}
&&{\cal E}_{n+1}={\cal E}_n + \frac{8}{\pi B}\sin\frac{\pi B}{4}\ \times\nonumber\\
&&  \cos\frac{\pi}{2}\left(\frac{4 B}{\beta^2}
(a_+ -\pi/2)-(n+1/2)B\right),\nonumber\\
\end{eqnarray}
which should be compared with (\ref{energyspectrum}). Before making the 
comparison
it is perhaps worth noting that as $\beta \rightarrow 0$
\begin{equation}
\rho_n\rightarrow (\pi -a_+),\ {\cal E}_n\rightarrow (n+1/2)\omega_0,
\end{equation}
where $\omega_0=2\sin a_+$ is the lowest possible breather frequency.

Comparing (\ref{WKBenergyspectrum}) with (\ref{energyspectrum}) for 
several values of $n$ yields two pieces of information. First of
all the mass of the sinh-Gordon particle must be given by
\begin{equation}
m(\beta)=\frac{8}{\pi B}\sin\frac{\pi B}{4},
\end{equation}
and, secondly, the parameter $E$ appearing in (\ref{shGordonR}) for
the reflection factor must be related to the boundary parameters and the 
bulk coupling by
\begin{equation}\label{Eexpression}
E(\beta, a_0,a_1)=(a_0+a_1)(1-B/2).
\end{equation}
This expression agrees perfectly with the classical limit as 
$\beta\rightarrow 0$, and with the result given by Ghoshal and 
Zamolodchikov for the Neumann boundary condition. There, $a_0=a_1=1/2$
and $E=1-B/2$ as they claimed. The expression (\ref{Eexpression}) also
agrees with  one loop perturbative calculations \cite{Cor98, Top97},
and with the known results at the special point $B=-2$ \cite{Ame95} 
(see \cite{Cor98}).

As far as the other parameter $F$ is concerned we know its classical limit,
and we also know its expansion to order $\beta^2$, at least to first order
in the difference $a_0-a_1$ (see \cite{Chen00}). Unfortunately, when the bulk
$Z_2$ symmetry is broken the perturbation calculations are substantially more
complicated, and indeed unfinished for arbitrary $a_0,a_1$ 
(see \cite{Chen00a}). These facts are, however, consistent
with the expression
\begin{equation}\label{Fexpression}
F(\beta, a_0, a_1)=(a_0-a_1)(1-B/2).
\end{equation}
In any case, had we chosen to replace $a_1$, say, by $-a_1$, we would
expect $E$ and $F$ to interchange (since $E$ by itself  is not invariant 
under such a change). 

Notice that since $0\le 1-B/2\le 1$, the assumptions 
we made concerning the regions in the parameter space for which the
reflection factor (\ref{shGordonR}) has poles are vindicated.

One slight criticism of the technique is the requirement on the boundary 
parameters to lie in the region where the boundary breathers exist. Otherwise,
strictly-speaking we could deduce nothing about the reflection factor since the
comparison we have utilised would not be valid. However, it is perhaps worth
noting that in the limit where $a_0=a_1\rightarrow 1/2$, the result  does agree 
with the earlier conjectures of Ghoshal and Zamolodchikov concerning the 
Neumann boundary conditions despite the obvious fact that the Neumann boundary
condition does not support boundary bound states.

\section{Weak-strong coupling duality}

We mentioned that the S-matrix describing the scattering of the sinh-Gordon 
particle is invariant under the exchange $\beta \rightarrow 4\pi/\beta$. It
is interesting to notice that there is a sense in which this duality extends to
the reflection factors also. Consider the triple of coupling constants
$(\beta, a_0, a_1)$. If we define a new triple by making the change
\begin{equation}
(\beta, a_0,a_1)\rightarrow \frac{4\pi}{\beta^2}(\beta, a_0, a_1),
\end{equation}
then it is simple to check that the reflection factor (\ref{shGordonR}) with 
$E$ and $F$ given by (\ref{Eexpression}) and (\ref{Fexpression}), respectively,
is invariant.

\section{Discussion and conclusions}

The sinh-Gordon model is just  about the simplest massive model one 
might contemplate beyond
free field theory. Yet it has been a long and tortuous road 
 to discover a dynamical argument which would enable us to derive 
a relationship
between the parameters occurring in the Lagrangian formulation of the model
and the parameters arising from other, purely algebraic considerations---in the sense of
deductions made using the boundary version of the Yang-Baxter equations together 
with the bootstrap. Our argument is certainly non-perturbative but it might not
be exact. We hope that it will turn out to be exact but we cannot be sure.
\footnote{
It should be remarked that Al. Zamolodchikov has also calculated (but not yet 
published) the parameter
dependence derived using quite different arguments \cite{Zam99}. }

The properties of the boundary breathers we have found are quite intriguing.
We should like to see an explanation for the surprising facts, for example
concerning the dependence of the classical action on the boundary parameters.
We would also be interested in exploring the situation with two boundaries.
There, the field theory is confined to an interval and the spectrum of states
within the interval should depend on the bulk coupling and four boundary 
parameters. One might expect the energy spectrum to be determined by
a pair of reflection factors, one for each boundary, but that idea requires
the assumption of 
factorization. In other words, if the boundaries behave independently we would
expect a relationship of the type
\begin{equation}
e^{4ikL}R^{(L)}(k)R^{(-L)}(-k)=1,
\end{equation}
where the two boundaries are situated at $x=\pm L$, and the corresponding 
reflection factors are $R^{(\pm L)}$. 
It would be interesting to see how this spectrum might be compatible 
with the quantization of periodic classical solutions.

There are several directions to go beyond sinh-Gordon. For example, the next
simplest model with a single scalar field is the model based on $a_2^{(2)}$
data. There, instead of having two boundary parameters, there is only one
but it can arise in two distinct ways \cite{Bow95}. For this reason, we 
expect a  greater variety of boundary breathers. The classical reflection 
factors are
known \cite{Cor96} but at the moment the analogue of Ghoshal's formula 
has not been found.
Indeed, analysing the boundary breathers might be of great assistance for
this case, to serve as a guide in finding the correct expressions. Besides, 
there is a lot to do to classify completely 
all the reflection data
for all the affine Toda models, extending what is already known for the 
$a_n^{(1)}$ series \cite{Del99a}.

\acknowledgments

I am grateful to my co-workers Medina Ablikim, Alireza Chenaghlou, Gustav Delius, Uli Harder,
Viktoria Malyshenko and
Anne
 Taormina for enjoyable collaborations, to the Universities of Lyon, Mons, 
Montpellier and Trieste for their hospitality at various times,  and to the European 
Commission for financial support under the Training and Mobility of Researchers 
Network Contract ERBFMRX-CT-960012.

\end{document}